\documentclass{article}

\usepackage{amsmath,amssymb,graphicx, hyperref, psfrag, epsfig, mathrsfs}
\usepackage{color}
\usepackage{cite}
\usepackage{hyperref}

\usepackage{amsthm}
\theoremstyle{definition}

\makeatletter
\@addtoreset{equation}{section}
\renewcommand{\theequation}{\thesection.\@arabic\c@equation}
\makeatother
\makeatletter
\renewcommand\appendix{\par
  \setcounter{section}{0}%
  \setcounter{subsection}{0}%
  \gdef\thesection{Appendix \@Alph\c@section }
  \renewcommand{\theequation}
  {\Alph{section}.\arabic{equation}}
}
\makeatother

\def \be {\begin{equation}}
\def \ee {\end{equation}}
\def \ba {\begin{array}}
\def \ea {\end{array}}
\def \bea{\begin{eqnarray}}
\def \eea{\end{eqnarray}}

\def \a {\alpha}

\def \l {\lambda}

\def \p {\partial}
\def \f {\frac}

\def \nn {\nonumber}

\def \ma {\mathcal}

\def \mc {\mathcal}

\def \lt {\left}
\def \rt {\right}

\def \ra {\rightarrow}

\def \sr {\sqrt}

\title{\textbf{Critical magnetic field in holographic superconductor in Gauss-Bonnet gravity with Born-Infeld electrodynamics}}
\author{
Sheng-liang Cui$^{}$\footnote{shlcui@pku.edu.cn},\,
Zhao Xue$^{}$\footnote{xuezhao2012@pku.edu.cn}
}

\date{}

\begin{document}

\maketitle

\begin{center}
{\it
$^{}$Department of Physics and State Key Laboratory of Nuclear Physics and Technology, Peking University, Beijing 100871, P.R. China\\

}
\vspace{10mm}
\end{center}

\begin{abstract}

In the paper using matching method in the probe limit, we investigate some properties of holographic superconductor in Gauss-Bonnet gravity with Born-Infeld electrodynamics . We discuss  the effects of the Gauss-Bonnet coupling $\a$ and Born-Infeld parameter $b$  on the critical temperature and
 condensate. We find that both of $\a$ and $b$ make the critical temperature decrease, which implies the condensate harder to form.
  Moreover we study the magnetic effect on holographic superconductor and obtain that the ratio between the critical magnetic field and the square of the critical temperature increases from zero as the temperature is lowered below the critical value $T_c$, which agrees well with the former results. We also find the critical magnetic field is indeed affected by  Gauss-Bonnet coupling, but not by Born-Infeld parameter.

\end{abstract}

\baselineskip 18pt
\thispagestyle{empty}

\newpage


\section{Introduction}

     The gauge/gravity correspondence\cite{MaldacenaRE}-\cite{WittenQJ}, which is the most remarkable achievement in string theory, states that string theory in asymptotic anti de Sitter  spcetimes can be dual to a conformal field theory on  the boundary, which is a realization of holographic theorem. It provides a powerful tool to  research strongly coupled field theory, such as quark confinement in QCD, cold atom system and high temperature superconductor in condensed matter physics.

A holographic model of superconductor in 2+1 dimensions was constructed in \cite{HHHone}, which gives that the gravitational duals to high temperature superconducting systems. The model consists of a black hole in an asymptotic AdS spacetimes coupled to a $U(1)$ gauge field and a minimally coupled complex scalar field.
Using the  gauge/gravity duality, the $U(1)$ gauge symmetry in the bulk correspond to global symmetry in the dual boundary conformal field theory.
When the temperature T is below the critical value $T_c$, the black hole will become unstable and grow scalar "hair", which corresponds to the  condensing of the operator dual to the scalar field. The condensate breaks  the $U(1)$ gauge symmetry in the bulk spontaneously, giving a mass to the gauge field. On the boundary side this process corresponds to  the non-trivial expectation value of the operator dual to scalar field, which is considered as a superconducting phase transition.

There are many works studying various holographic superconductor models \cite{Horowitz:2008bn}-\cite{Banerjee:2012vk}, for reviews see \cite{Hartnoll:2008kx,Hartnoll:2009sz,Horowitz:2010gk}, among which some effects such as high curvature corrections by changing the Einstein gravity to Gauss-Bonnet gravity, high electric-magnetic field terms by introducing Born-Infeld electrodynamics and external magnetic field etc have been discussed. The effect of high curvature corrections on the holographic superconductors in Gauss-Bonnet gravity has been analyzed numerically and analytically by Sturm-Liouville eigenvalue problem method \cite{Banerjee:2012vk}-\cite{Ge:2010aa} and by the matching  method \cite{Gregory:2009fj}\cite{Pan:2009xa}. They found that near the critical temperature, the critical exponent of the condensate is $\f{1}{2}$, and as the Gauss-Bonnet coefficients grows, the critical temperature decreases, so it makes the condensate harder to form.
The superconductors with Born-Infeld electrodynamics have been investigated in \cite{Gangopadhyay:2012am}\cite{Gangopadhyay:2012np}, in which they found that Born-Infeld parameter makes the condensate harder to form.  There are some works about the effects of magnetic field on holographic superconductor in Einstein gravity \cite{Nakano:2008xc}-\cite{Roychowdhury:2012hp}. In \cite{Roychowdhury:2012hp}, the authors studied the effects of magnetic field on the  holographic superconductor in presence of nonlinear  corrections to the gauge field. They managed to obtain the critical exponent $\f{1}{2}$ and the upper critical magnetic field formula.

In spite of those works, there are some issues aren't yet to be explored.  it was noticed that all the above analysis are most performed with the high curvature corrections or the high electric-magnetic field term corrections singly, questions that naturally arise are that what will happen if we involve simultaneously the high curvature corrections and the high electric-magnetic field term corrections, how will the two parameters $\alpha$ and $b$ affect the behavior of holographic superconductor in the presence of an external magnetic field and whether there is correlation between two parameters $\alpha$ and $b$,  In order to address the above mentioned issues,
in this paper, we extended to investigate the holographic superconductor in Gauss-Bonnet gravity with Born-Infeld electrodynamics by using the analytic method in \cite{Gregory:2009fj, Ge:2010aa, Roychowdhury:2012hp} . We find that both of the Born-Infeld parameter $b$ and the Gauss-Bonnet coupling $\alpha$ make the critical temperature $T_c$ decrease, so the scalar condensate becomes harder. In the presence of an external magnetic field,  we obtain that the critical magnetic field value increase as temperature is lowered below the critical temperature $T_c$. In the region of $T\sim T_c$ , the formula of critical magnetic field agrees well with the standard expression \cite{poole2007}.

The paper is organized as follows. In section 2, we will introduce the model of superconductor in the background (4+1) dimensional Gauss-Bonnet-AdS planar black hole with Born-Infeld electrodynamics in the probe limit by the matching method, and analyze the relations of the critical temperature $T_c$, the Born-Infeld parameter $b$ and the Gauss-Bonnet coupling $\alpha$.  In section 3, we will involve  external magnetic filed into the model and study the effect of magnetic filed on holographic superconducting phase transition. Conclusions will be drawn in section 4.

\section{Superconductor in Gauss-Bonnet gravity with Born-Infeld electrodynamics }\label{s2}

In this section we consider the holographic superconductor in Gauss-Bonnet gravity with Born-Infeld electrodynamics following the method developed in \cite{Gregory:2009fj}. The action of Gauss-Bonnet gravity in a 5-dimensional space-time can be written as
\begin{equation}
\mathcal{S}=\int d^{5}x \sqrt{-g}\left[R+\frac{12}{L^{2}}+\frac{\alpha}{2}(R^{2}-4 R^{\mu\nu}R_{\mu\nu}+R^{\mu\nu\rho\sigma}R_{\mu\nu\rho\sigma})\right]. \label{eq1}
\end{equation}
The Lagrangian density is
\be
\mathcal{L}=\frac{1}{b}\left(1-\sqrt{1+\frac{b}{2}F^{\mu\nu}F_{\mu\nu}} \right)-|\nabla_{\mu}\psi- iA_{\mu}\psi |^{2}-m^{2}\psi^{2}.\nn
\ee
Here $\alpha $ is the Gauss-Bonnet coupling and $ b $ is the Born-Infeld parameter. For convenience, we set $L=1$ in the later discussion.
The gravity background solution for Gauss-Bonnet action is a planar black hole in the probe limit, which be described by the following metric
\be
ds^{2}=-f(r)dt^{2}+f^{-1}(r)dr^{2}+r^{2}(dx^{2}+dy^{2}+dz^{2}),
\ee
where \be f(r)=\frac{r^{2}}{2\alpha}\left(1-\sqrt{1-4\alpha\left(1-\frac{M}{r^{4}}\right) } \right).\ee
One can define an effective radius of the AdS spacetime by
\be L_{eff}^2=\f{2\alpha}{1-\sqrt{1-4\alpha}}. \ee
The Hawking temperature of the black hole is given by
\be
T=\frac{f'(r)}{4\pi}|_{r=r_+}=\frac{r_+}{\pi}, \label{bhT}
\ee
where $r_+$ is the horizon radius of the black hole.
Taking a static ansatz,
\be A_{\mu}=(\phi(r),0,0,0),\;\;\;\;\psi=\psi(r). \ee
The equations of motion for the complex scalar field $\psi$ and the Maxwell field $A_\mu$ are
\begin{equation}
\partial_{r}^{2}\phi + \frac{3}{r}\lt(1-b(\partial_{r} \phi)^{2}\rt)\partial_{r}\phi -\frac{2\psi^{2}\phi}{f}\lt(1-b(\partial_{r}\phi)^{2}\rt)^{3/2}=0\label{eq2}
\end{equation}
and
\begin{equation}
\partial_{r}^{2}\psi + \left(\frac{3}{r}+\frac{\partial_{r}f}{f} \right)\partial_{r}\psi + \frac{\phi^{2}\psi}{f^{2}}-\frac{m^{2}\psi}{f}=0\label{eq3}.
\end{equation}
We change the variable and set
\be z=\frac{r_+}{r}, \ee
so the equations of motion of $\psi$ and $\phi$ become
\be
\phi''(z)-\frac{1}{z} \phi'(z)+\frac{3bz^3}{r_+^2}\phi'^3(z)-\frac{2r_+^2\psi^2\phi}{z^4 f(z)}\left(1-\frac{bz^4}{r_+^2}\phi'^2(z)\right)^\frac{3}{2}=0 \label{eom1}
\ee
and
\be
\psi''(z)-\frac{1}{z}\psi'(z)+\f{f'(z)}{f(z)}\psi'(z)+\f{r_+^2\phi^2}{z^4 f^2(z)}\psi-\f{r_+^2 m^2}{z^4 f(z)}\psi=0\label{eom2}.
\ee
where a prime denotes the derivative with respective to $z$ .

We choose the mass of the scalar field to be $m^2=-\f{3}{L_{eff}^2}$, which is above the BF bound. At the horizon ($z=1$) for regularity , we take the bound conditions,
\be
\phi(1)=0 \label{bc1} .
\ee
From (\ref{eom2}) and (\ref{bc1}), we obtain
\be \psi(1)=\f{4}{3}\psi'(1) \label{bc11}. \ee
 In the asymptotic region ($z\to 0$ ),  the solutions behave as
\be \phi(z)=\mu-\f{\rho}{r_+^2}z ^2\ee
and
\be \psi(z)=\psi_-\f{z}{r_+}+\psi_+\f{z^3}{r_+^3}. \ee
Here, $\mu$ and $\rho$ are interpreted as a chemical potential and charge density respectively.
We choose $\psi_+=\langle \ma{O} \rangle $ and set $\psi_-=0$, because $\psi_-$ is non-renormalizable, which corresponds to sources.
We expand both $\phi$ and $\psi$ near the horizon of the black hole $z=1$,
\be
\phi(z)=\phi(1)+\phi'(1)(z-1)+\f{1}{2}\phi''(1)(z-1)^2+\cdots  \label{phie}
\ee
\be
\psi(z)=\psi(1)+\psi'(1)(z-1)+\f{1}{2}\psi''(1)(z-1)^2+\cdots \label{psie}
\ee

From the equations of motion (\ref{eom1}) and  (\ref{eom2}), using the boundary conditions (\ref{bc1}) and  (\ref{bc11}), we obtain
\be
\phi''(1)=\phi'(1)-\f{3b\phi'^3(1)}{r_+^2}-\f{\phi'(1)\psi^2(1)}{2}\lt(1-\f{3b}{2r_+^2}\phi'^2(1)\rt)+\mathcal{O}(b^{2}) \label{phi1}
\ee
and
\be
\psi''(1)=(-\f{15}{32}+3\alpha)\psi(1)-\f{\phi'^2(1)}{32r_+^2}\psi(1)\label{psi1}+\mathcal{O}(b^{2}).
\ee

Thus we obtain $\phi(z)$ and $\psi(z)$  near $z=1$ from (\ref{phie}) and (\ref{psie})
\begin{eqnarray}
\phi(z)&=&\phi'(1)(z-1)+\f{\phi'(1)}{2}\lt[1-\f{3b\phi'^2(1)}{r_+^2}-\f{\psi^2(1)}{2}\lt(1-\f{3b\phi'^2(1)}{2r_+^2}\rt)\rt](1-z)^2 \nn \\
&&+\mathcal{O}\left((1-z)^{3}\right),\\
\psi(z)&=&\psi(1)(\f{1}{4}+\f{3z}{4})+\f{\psi(1)}{2}\lt[(3\alpha-\f{15}{32})-\f{\phi^2(1)}{32r_+^2}\rt](1-z)^2+\mathcal{O}\left((1-z)^{3}\right).
\end{eqnarray}
where  all calculations were done to the first order in the Born-Infeld parameter $b$.

In order to match these two sets of asymptotic solutions at some intermediate point $z=z_m$ , the following four equations must be satisfied \cite{Gregory:2009fj},
\bea
&&\mu-\f{\rho}{r_+^2}z_m^2 = \phi'(1)(z_m-1)+\f{1}{2}[\phi'(1)-\f{3b\phi'^3(1)}{r_+^2}-\f{\phi'(1)\psi^2(1)}{2}(1-\f{3b\phi'^2(1)}{2r_+^2})](z_m-1)^2 , \label{ma1}\\
&&-\f{2\rho}{r_+^2}z_m = \phi'(1)+[\phi'(1)-\f{3b\phi'^3(1)}{r_+^2}-\f{\phi'(1)\psi^2(1)}{2}(1-\f{3b\phi'^2(1)}{2r_+^2})](z_m-1) \label{ma2}, \\
&&\f{z_m^3}{r_+^3}\psi_+ = \f{1}{4}\psi(1)+\f{3}{4}\psi(1)z_m+\f{1}{2}\psi(1)[(3\alpha-\f{15}{32})-\f{\phi'^2(1)}{32r_+^2}](z_m-1)^2 ,\label{ma3} \\
&&\f{3z_m^2}{r_+^3}\psi_+ = \f{3}{4}\psi(1)+\psi(1)[(3\alpha-\f{15}{32})-\f{\phi'^2(1)}{32r_+^2}](z_m-1)\label{ma4}.
\eea
We set $\beta=\p_r\phi(r_+)=-\f{\phi'(1)}{r_+}$. From (\ref{ma2}) we obtain
\be
\psi^2(1)=\f{2}{z_m-1}\lt[z_m-\f{2\rho z_m}{\beta r_+^3}-\f{3b\rho\beta z_m}{r_+^3}+3b\beta^2(1-\f{z_m}{2})\rt]+ \mc{O}(b^2).
\ee
Using temperature formula (\ref{bhT}),  we have
\be
\psi^2(1)=\f{2\lt[z_m+3b\beta^2(1-\f{1}{2}z_m)\rt]}{1-z_m}(\f{T_c^3}{T^3})(1-\f{T^3}{T_c^3})+\mc{O}(b^2) \label{psi11},
\ee
where we have defined $T_c$ as, 
\be
T_c=\f{(2\rho)^\f{1}{3}}{\pi\beta^\f{1}{3}}(1-3b\beta^2\f{1-z_m}{z_m})^\f{1}{3} \label{Tc}.
\ee
For $T\sim T_c$, using (\ref{psi11}) we have
\be
\psi(1)=\sqrt{\f{6\lt[z_m+3b\beta^2(1-\f{1}{2}z_m)\rt]}{1-z_m}} \sr{1-\f{T}{T_c}}+\mc{O}(b^2). \label{condensate}
\ee
From  (\ref{ma3}) and  (\ref{ma4}) , we obtain
\be
\psi_+=\f{5+3z_m}{4z_m^2(3-z_m)}r_+^3\psi(1) , \label{psi+}
\ee
and
\be
\beta^2=96\alpha-15+\f{48(1+2z_m)}{(1-z_m)(3-z_m)}.
\ee
For $T \sim T_c$, we obtain the expectation value $\langle \ma{O} \rangle$ from (\ref{condensate}) and (\ref{psi+})
\be
\f{\langle \ma{O} \rangle}{ T_c^3} =\psi_+=\pi^3\f{5+3z_m}{4z_m^2(3-z_m)}\sqrt{\f{6\lt[z_m+3b\beta^2(1-\f{1}{2}z_m)\rt]}{1-z_m}}\sr{1-\f{T}{T_c}}+\mc{O}(b^2),
\ee
where we have normalized by the critical temperature to obtain a dimensionless quantity.

\begin{figure}[ht]
 \centering
 \includegraphics[width=0.6\textwidth]{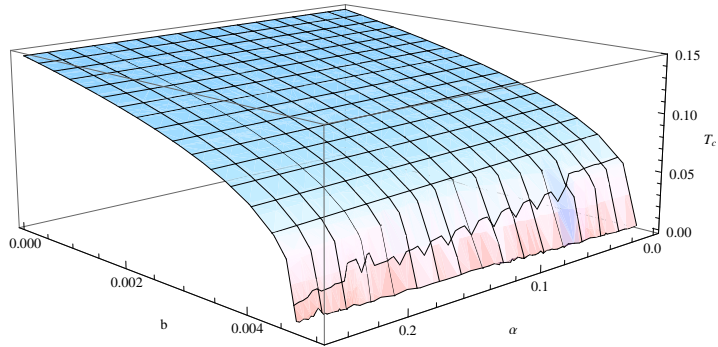}
 \caption{The critical temperature $T_c$ as a function of $\alpha$ and $b$.}
 \label{T}
\end{figure}

\begin{figure}[h]
 \centering
 \includegraphics[width=0.45\textwidth]{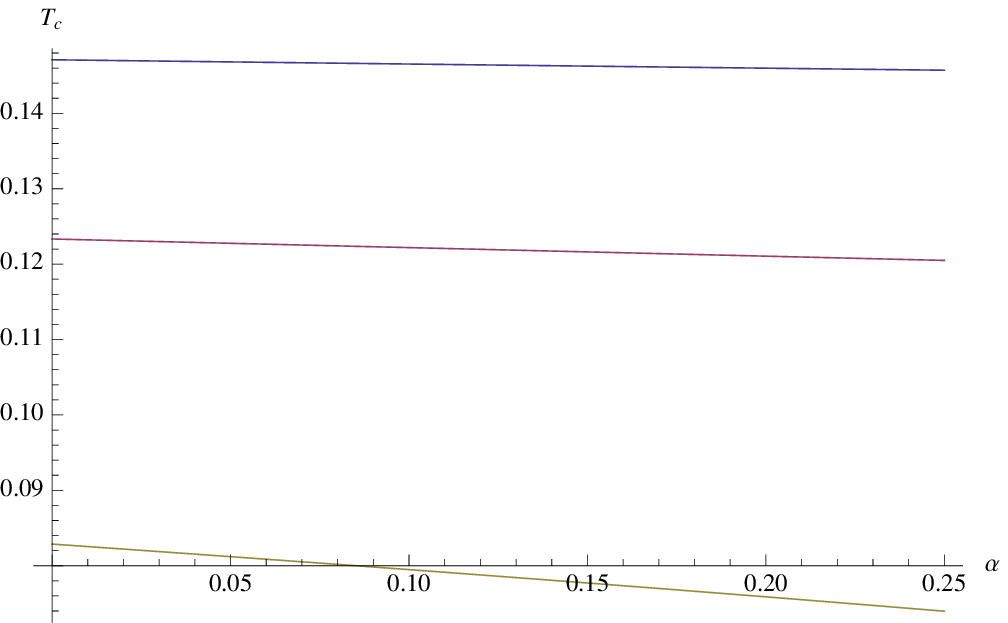}$\qquad\qquad$
 \includegraphics[width=0.45\textwidth]{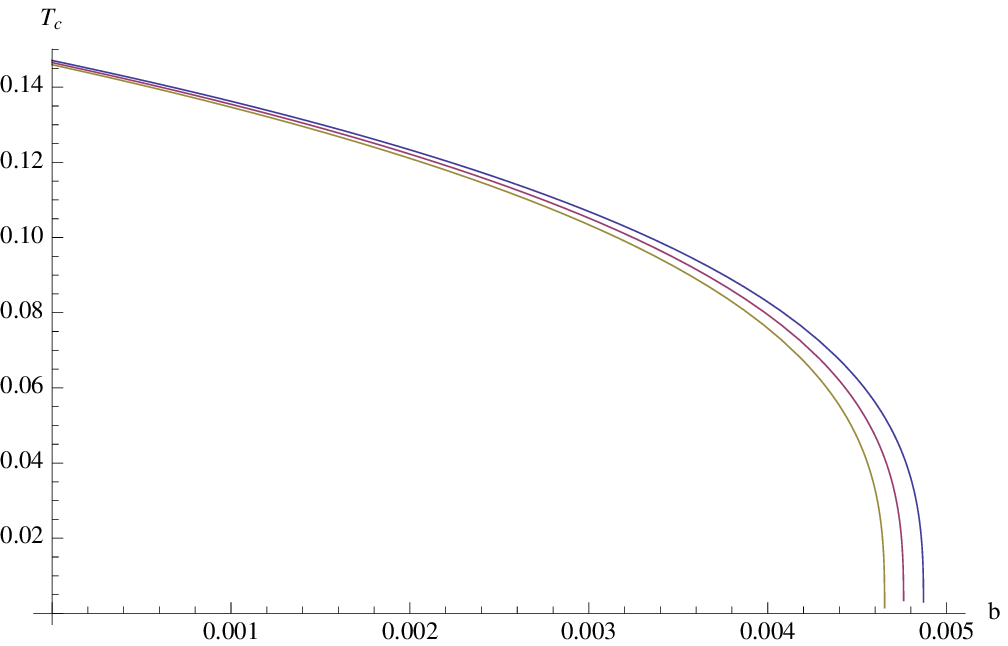}
   \caption{$T_c$ varies with $\alpha$ for various values of $b=0, 0.002, 0.004$ from top to bottom on the left. $T_c$ varies with $b$ for various of $\alpha=0, 0.1, 0.2$ from top to bottom on the right. Here choose $\rho=1$ and $z_m=\f{6}{7}$ (which be accounted for in section 3).}
 \label{Tab}
\end{figure}

\begin{figure}[h]
 \centering
 \includegraphics[width=0.7\textwidth]{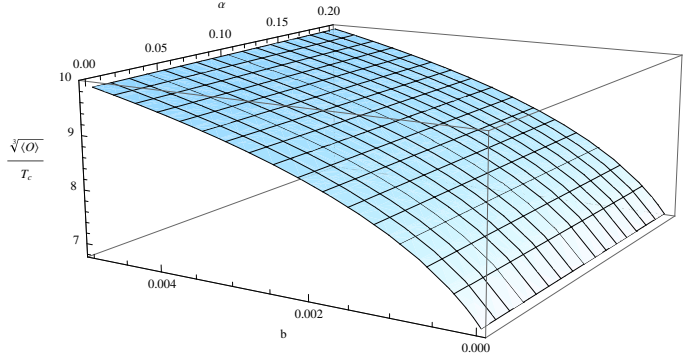}
 \caption{The condensate as a function of $\alpha$ and $b$.}
 \ \label{O}
\end{figure}

\begin{figure}[h]
 \centering
 \includegraphics[width=0.45\textwidth]{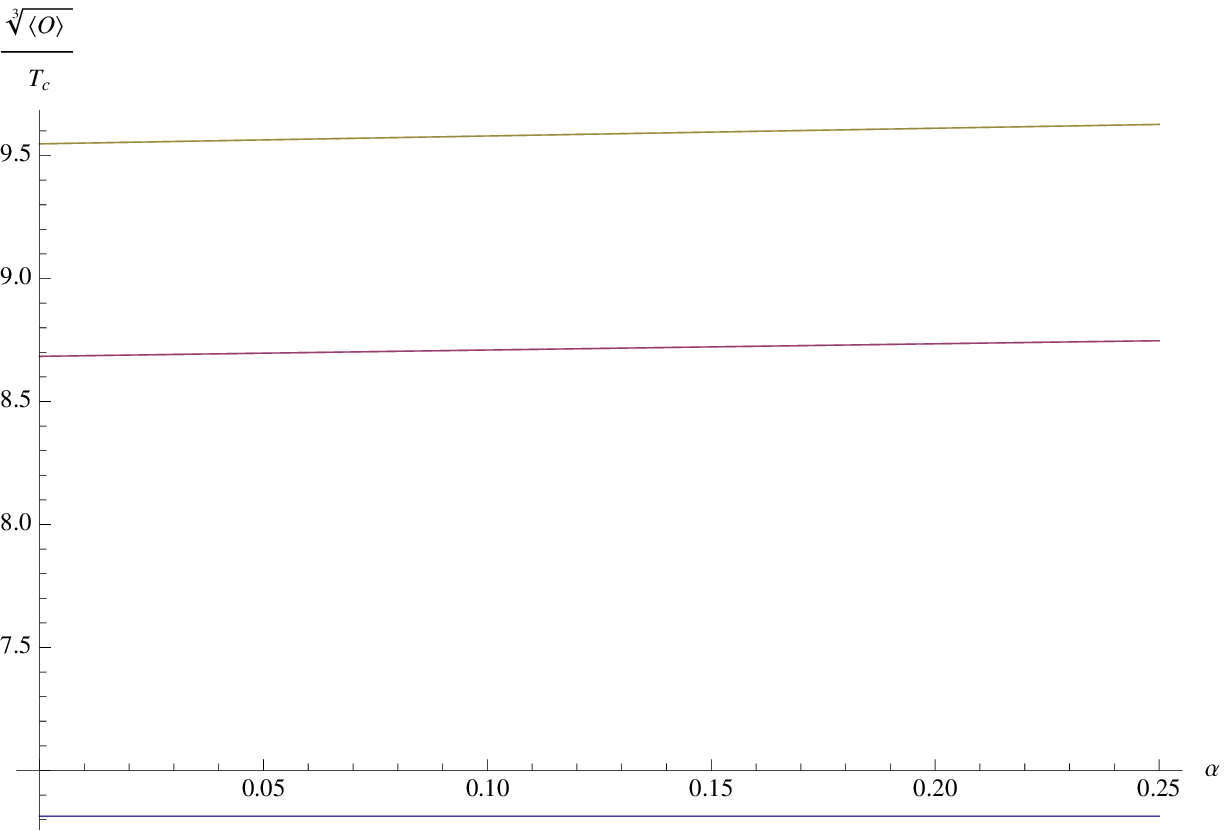}$\qquad\qquad$
 \includegraphics[width=0.45\textwidth]{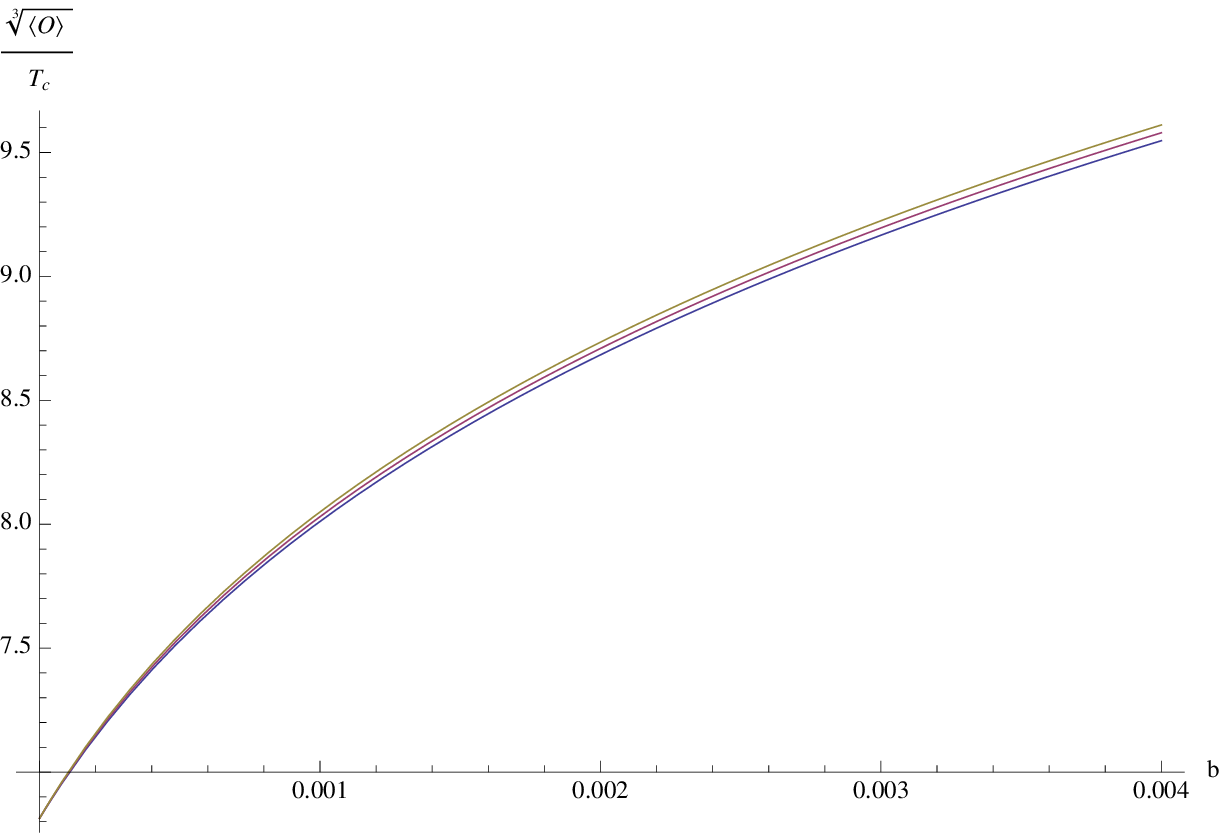}
   \caption{The condensate $\langle \ma{O} \rangle $ varies with $\alpha$ for various of $b=0, 0.002, 0.004$ from bottom to top on the left. The condensate $\langle \ma{O} \rangle $ varies with $b$ for various of $\alpha=0, 0.1, 0.2$ from bottom to top on the right. Here choose $z_m=\f{6}{7}$ and $\rho=1$.}
 \label{Oab}
\end{figure}

We find $\langle \ma{O} \rangle \propto\sr{1-\f{T}{T_c}}$ near the critical temperature, so the critical exponent of the condensate is $\f{1}{2}$, that is typical of the second order phase transitions. From (\ref{Tc}), we find the critical temperature $T_c$ changes with the Gauss-Bonnet coupling $\alpha$ and the Born-Infeld parameter $b$. This results are shown in figure \ref{T} and figure \ref{Tab} . From figure \ref{Tab}, we see that Gauss-Bonnet term makes the critical temperature a little lower, but the critical the temperature visibly decreases as Born-Infeld parameter $b$ increasing. So the increasing of $\alpha$ and $b$ both make the superconducting phase transition harder. In figure \ref{O} and \ref{Oab}, we have plotted the behavior of the condensate for different $\alpha$ and $b$.  We find that the ratios between the condensate value and the cube of the critical temperature $\f{\langle \ma{O} \rangle}{T_c^3}$ become larger as the $\alpha$ or $b$ increases. These results agree with the conclusions \cite{Gregory:2009fj, Gangopadhyay:2012np}.

\section{Magnetic field effect on holographic superconductor }\label{s3}

In this section we study the effect of an external static magnetic field on the superconductor. According to the dictionary of the gauge/gravity duality, magnetic field on the boundary is the asymptotic value of the magnetic field in the bulk,  $B({\bf{x}})=F_{xy}({\bf{x}}, z\ra 0)$.
We take the ansatz, 
\be
A=(A_t,0,A_y,0,0),~~~where~~~  A_t=\phi(z),~~~A_y=Bx,~~~\psi=\psi(x,z),
\ee
where we take a constant magnetic field $B$ for convenience.
The equation of motion for the scalar $\psi$ is
\bea
\psi''(x,z)&+&\f{f'(z)}{f(z)}\psi'(x,z)-\f{1}{z}\psi'(x,z)+\f{r_+^2\phi^2(z)}{z^4f^2(z)}\psi(x,z)+\f{3r_+^2}{z^4f(z)}\psi(x,z) \nn\\
&+&\f{1}{z^2f(z)}\lt(\p_x^2\psi(x,z)-B^2x^2\psi(x,z)\rt)=0.
\eea
We set $\psi(x,z)=R(z)X(x)$,
\be
z^2f(z)\lt[\f{R''}{R}+\f{f'R'}{fR}-\f{R'}{zR}+\f{r_+^2\phi^2(z)}{z^4f^2(z)}+\f{3r_+^2}{z^4f(z)}\rt]+\lt[\f{\p_x^2 X}{X}-B^2x^2\rt]=0.
\ee
From \cite{Albash:2008eh} we know that this is a eigenvalues equation,
\be
-\p_x^2X(x)+B^2x^2X(x)=\lambda_nBX(x),
\ee
where $\l_n=2n+1$ , $n$ for integer. We take $n=0$
\be
R''(z)+\f{f'(z)}{f(z)}R'(z)-\f{1}{z}R'(z)+\f{r_+^2\phi^2(z)}{z^4f^2(z)}R(z)+\f{3r_+^2}{z^4f(z)}R(z)=\f{BR(z)}{z^2f(z)}. \label{eomB}
\ee

At the horizon ($z=1$), for regularity we take
\be
\phi(1)=0.
\ee
From (\ref{eomB}) we obtain
\be R'(1)=(\f{3}{4}-\f{B}{4r_+^2})R(1) \label{bc2} \ee
and
\be
R''(1)=\lt(\f{B^2}{r_+^4}+\f{2B-32\alpha B-\phi'^2(1)}{r_+^2}+96\alpha-15)\rt)\f{R(1)}{32}.
\ee
We expand $R(z)$ near the horizon of the black hole $(z=1)$
\be
R(z)=\phi(1)+R'(1)(z-1)+\f{1}{2}R''(1)(z-1)^2+\cdots  \label{Ze}
\ee

So near the horizon $z\ra1$, we have
\be
R(z)=\lt[1-(1-z)\lt(\f{3}{4}-\f{B}{4r_+^2}\rt)+\f{(1-z)^2}{64}\lt(\f{B^2}{r_+^4}+\f{2B-32\alpha B-\phi'^2(1)}{r_+^2}+96\alpha-15\rt)\rt]R(1).
\ee
In the asymptotic region($z\ra0$), the solution behaves as
\be R(z)=R_-\f{z}{r_+}+R_+\f{z^3}{r_+^3}. \ee
We set $R_+=\langle \ma{O} \rangle $ and $R_-=0$ .
To match these two sets of asymptotic solutions at an intermediate point $z=z_m$, we require the following equations:
\bea
\f{R_-}{r_+^3}z_m^3& =&\lt[1-(1-z_m)\lt(\f{3}{4}-\f{B}{4r_+^2}\rt)\rt]R(1) \nn \\
&&+\lt[\f{(1-z_m)^2}{64}\lt(\f{B^2}{r_+^4}+\f{2B-32\alpha B-\phi'^2(1)}{r_+^2}+96\alpha-15\rt)\rt]R(1),\\
3\f{R_-}{r_+^3}z_m^2 &=&\lt[(\f{3}{4}-\f{B}{4r_+^2})-\f{(1-z_m)}{32}\lt(\f{B^2}{r_+^4}+\f{2B-32\alpha B-\phi'^2(1)}{r_+^2}+96\alpha-15\rt)\rt]R(1).
\eea
From above equations, we obtain
\be
\f{B^2}{r_+^4}+\lt(2-32\alpha+\f{16(3-2z_m)}{(1-z_m)(3-z_m)}\rt)\f{B}{r_+^2}+\lt(-\f{\phi'^2(1)}{r_+^2}+96\alpha-15+\f{48(1+2z_m)}{(1-z_m)(3-z_m)}\rt)=0,
\ee
which has a solution
\bea
\f{B}{r_+^2}&=&\sqrt{\lt[1-16\alpha+\f{8(3-2z_m)}{(1-z_m)(3-z_m)}\rt]^2-\lt[-\f{\phi'^2(1)}{r_+^2}+96\alpha-15+\f{48(1+2z_m)}{(1-z_m)(3-z_m)}\rt]} \nn \\
&&-\lt[1-16\alpha+\f{8(3-2z_m)}{(1-z_m)(3-z_m)}\rt].
\eea

When $B\sim B_c$ , $\psi\sim0$, so equation of motion for $\phi$ from (\ref{eom1}) is
 \be
\phi''(z)-\frac{1}{z} \phi'(z)+\frac{3bz^3}{r_+^2}\phi'^3(z)+\mathcal{O}(b^{2})=0. \label{eom11}
\ee
There is an approximate solution upto the first order of Born-Infeld parameter $b$, which match the boundary condition for $\phi$ at $z\ra 0$ and  at $z\ra 1$,
\be
\phi(z)=\f{\rho}{r_+^2}(1-z^2+\f{b\rho^2 z^8}{2r_+^6}-\f{b\rho^2}{2r_+^6})+\mathcal{O}(b^{2}),
\ee
where $\rho$ is charge density. 
So
\be
\phi'(1)=-\f{2\rho}{r_+^2}(1-\f{2b\rho^2}{r_+^6})+\mathcal{O}(b^2) \label{phi1}.
\ee

Substitute (\ref{phi1}) into $B_c$,
\bea
B_c&=&\pi^2\beta \gamma^3\f{T_c^3}{T}\Bigg[\sqrt{(1+\f{16\lt[1-8\alpha+16\alpha^2-\f{8z_m+16\alpha(3-2z_m)}{(1-z_m)(3-z_m)}+\f{4(3-2z_m)^2}{(1-z_m)^2(3-z_m)^2}\rt]}{\beta^2}\lt(\f{T}{\gamma T_c}\rt)^6}  \nn \\
&&-\f{1-16\alpha+\f{8(3-2z_m)}{(1-z_m)(3-z_m)}}{\beta}\lt(\f{T}{\gamma T_c}^3\rt)\Bigg], \label{B1}
\eea
where
 \be
 \gamma=\lt(1-\f{1}{2}b\beta^2\lt(\f{Tc}{T}\rt)^6+3b\beta^2\f{1-z_m}{z_m}\rt)^\f{1}{3},
 \ee
\be
\beta=\sqrt{96\alpha-15+\f{48(1+2z_m)}{(1-z_m)(3-z_m)}} .
\ee
$T_c$ is the critical temperature (\ref{Tc}) in absent of an external magnetic field in section II.
From (\ref{B1}), $B_c=0$ ~at $T=\gamma T_c$. But we know the critical temperature (\ref{Tc}) in absent of an external magnetic field in section II.
 If and only if $z_m=\f{6}{7}$, $\gamma=1$, that is $T=T_c$, $B=0$ matches the result (\ref{Tc}).
 So the external magnetic field on the superconductor is
 \bea
\f{B_c}{ T_c^2}&=&\pi^2\beta\Bigg[\sqrt{(1+\f{16\lt[1-8\alpha+16\alpha^2-\f{8z_m+16\alpha(3-2z_m)}{(1-z_m)(3-z_m)}+\f{4(3-2z_m)^2}{(1-z_m)^2(3-z_m)^2}\rt]}{\beta^2}\lt(\f{T}{T_c}\rt)^6}  \nn\\
&&-\f{1-16\alpha+\f{8(3-2z_m)}{(1-z_m)(3-z_m)}}{\beta}\lt(\f{T}{T_c}\rt)^3\Bigg].\label{BTa}
\eea
where where we have normalized by the square of the critical temperature to obtain a dimensionless quantity.
 From the above result, we find that
  \be
 B_c \propto (1-\f{T}{T_c}),~~~~~~\text{at}~~ T \sim T_c,
 \ee
 which agrees with the parabolic law
 \be
 B_c(T)=B_c(0)[1-(\f{T}{T_c})^2].
 \ee

\begin{figure}[h]
 \centering
 \includegraphics[width=0.45\textwidth]{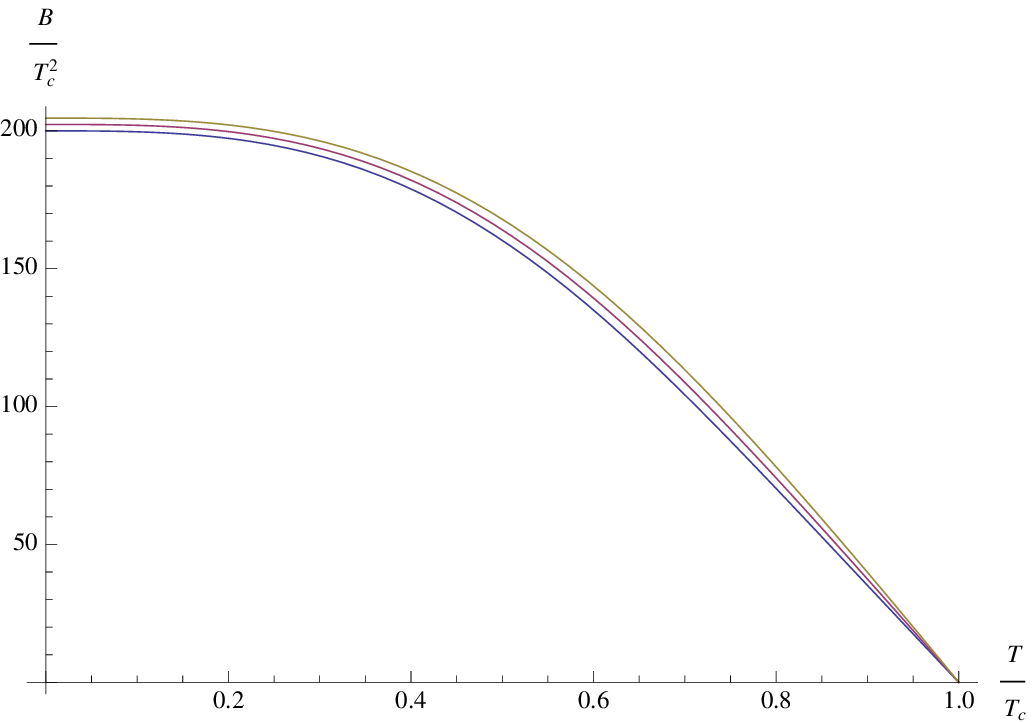}\qquad
 \includegraphics[width=0.45\textwidth]{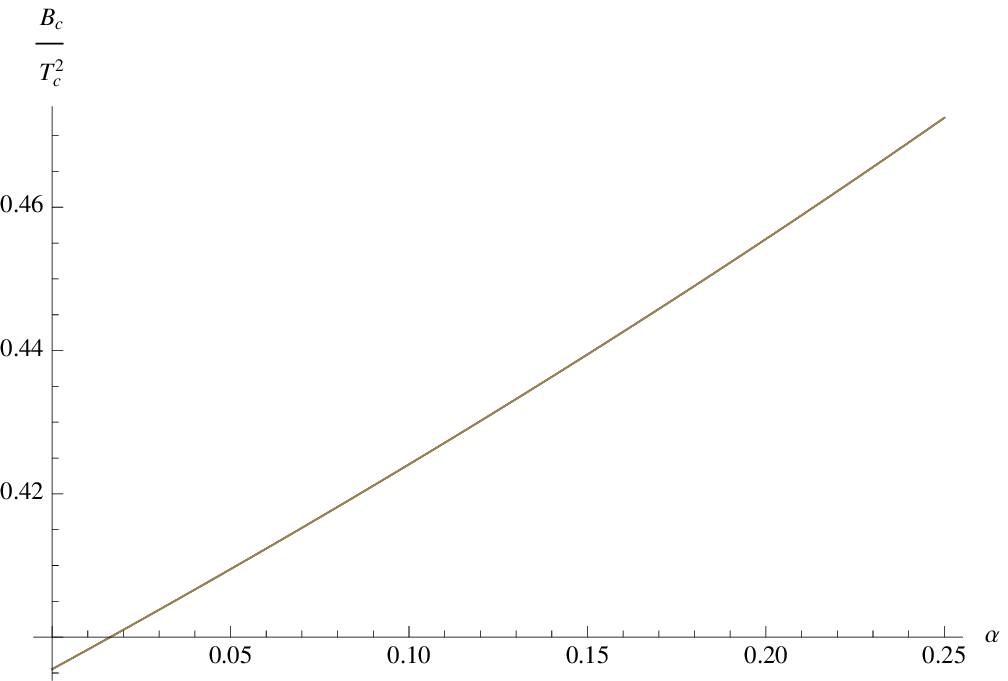}
   \caption{The magnetic field $\f{B_c}{T_c^2}$ varies with $\f{T}{T_c}$ for different $\alpha=0, 0.1, 0.2$ from bottom to top on the left.The critical magnetic field $\f{B_c}{T_c^2}$  varies with $\alpha$ on the right at $\f{T}{T_c}$=0.99.  }
 \label{BTab}
\end{figure}

It is noticed that the ratio between the upper magnetic field and the square of critical temperature is independent of the Born-Infeld parameter in the $T \sim T_c$. We show the result (\ref{BTa}) in Figure \ref{BTab}. From Figure \ref{BTab}, we see the critical magnetic fields is zero at $T=T_c$ and increase as the temperature decreases below the critical value $T_c$, which is in agreement with \cite{poole2007}. In the region $T \sim T_c$,  the result is credible, but as $T \to 0$, $B_c$ is divergent which is incredible because the back reactions isn't ignored. The critical magnetic field becomes larger as the Gauss-Bonnet coupling increases, this means that the  Gauss-Bonnet coupling makes diamagnetic ability stronger, which makes the phase transition of superconductor harder to happen.

\section{Conclusion}\label{s4}

In this paper we have investigated the holographic superconductor in Gauss-Bonnet gravity with Born-Infeld electrodynamics in an external magnetic field by using the matching method developed in \cite{Gregory:2009fj}. We found that the Gauss-Bonnet coupling $\alpha$ and the  Born-Infeld parameter $b$ both make the critical temperature $T_c$ decrease.  we also found that the exponent  of condensates of the operator is $\f{1}{2}$, and the ratio $\f{\langle \ma{O} \rangle}{T_c^3}$ between the condensate value and the cube of the critical temperature decreases when the Gauss-Bonnet coupling $\alpha$ or the  Born-Infeld parameter $b$ increases, which means $\alpha$ and $b$ both make the condensate harder. Those statements are agreement with the previous conclusions. But there is an ambiguity in the choice of the matching point $z_m$ . Surprisingly, this problem is solved in the process of studying the effect of external magnetic field on the holographic superconductor in Gauss-Bonnet gravity with Born-Infeld electrodynamics. In our model, the intermediate point $z_m$ must be $\f{6}{7}$ .We obtain that the critical magnetic field value increases as temperature is lowered below the critical temperature $T_c$. In spite of the divergence of the magnetic field as $T\ra 0$ , the result for $T\sim T_c$ agrees well with the known expression for the upper critical field strength \cite{poole2007}. The ratio $\f{B_c}{T_c^2}$ between the critical magnetic field and the square of the critical temperature  increases as the Gauss-Bonnet coupling $\alpha$ grows bigger.

\vspace*{10mm}
\noindent {\large{\bf Acknowledgments}}\\
 We would like to thank Bin Chen, Wei-shui Xu, Nan Bai, Bu Guan, Dan Wang and Xiao-bao Xu for their discussions and comments.
 \vspace*{5mm}

\providecommand{\href}[2]{#2}\begingroup\raggedright\endgroup






\begin{thebibliography}{10}
\bibitem{MaldacenaRE}
  J.~M.~Maldacena,
  Adv.\ Theor.\ Math.\ Phys.\  {\bf 2}, 231 (1998)
  [Int.\ J.\ Theor.\ Phys.\  {\bf 38}, 1113 (1999)]
  [arXiv:hep-th/9711200].

\bibitem{GubserBC}
  S.~S.~Gubser, I.~R.~Klebanov and A.~M.~Polyakov,
  Phys.\ Lett.\  B {\bf 428}, 105 (1998)
  [arXiv:hep-th/9802109].

\bibitem{WittenQJ}
  E.~Witten,
  Adv.\ Theor.\ Math.\ Phys.\  {\bf 2}, 253 (1998)
  [arXiv:hep-th/9802150].

\bibitem{HHHone}
  S.~A.~Hartnoll, C.~P.~Herzog and G.~T.~Horowitz,
  Phys.\ Rev.\ Lett.\  {\bf 101} (2008) 031601
  [arXiv:0803.3295 [hep-th]].

\bibitem{Hartnoll:2008kx}
  S.~A.~Hartnoll, C.~P.~Herzog and G.~T.~Horowitz,
  JHEP {\bf 0812} (2008) 015  [arXiv:0810.1563 [hep-th]].

\bibitem{Hartnoll:2009sz}
  S.~A.~Hartnoll,
   Class.\ Quant.\ Grav.\  {\bf 26} (2009) 224002  [arXiv:0903.3246 [hep-th]].  

\bibitem{Horowitz:2010gk}
  G.~T.~Horowitz,
 Lect.\ Notes Phys.\  {\bf 828} (2011) 313  [arXiv:1002.1722 [hep-th]].  

\bibitem{Horowitz:2008bn}
  G.~T.~Horowitz and M.~M.~Roberts,
  Phys.\ Rev.\ D {\bf 78} (2008) 126008
  [arXiv:0810.1077 [hep-th]].


\bibitem{Gregory:2009fj}
  R.~Gregory, S.~Kanno and J.~Soda,
  JHEP {\bf 0910} (2009) 010
  [arXiv:0907.3203 [hep-th]].

\bibitem{Pan:2009xa}
  Q.~Pan, B.~Wang, E.~Papantonopoulos, J.~Oliveira and A.~B.~Pavan,
  Phys.\ Rev.\ D {\bf 81} (2010) 106007
  [arXiv:0912.2475 [hep-th]].

\bibitem{Ge:2010aa}
  X.~-H.~Ge, B.~Wang, S.~-F.~Wu and G.~-H.~Yang,
  JHEP {\bf 1008} (2010) 108
  [arXiv:1002.4901 [hep-th]].

\bibitem{Pan:2010at}
  Q.~Pan and B.~Wang,
  Phys.\ Lett.\ B {\bf 693} (2010) 159
  [arXiv:1005.4743 [hep-th]].

\bibitem{Cai:2010cv}
  R.~-G.~Cai, Z.~-Y.~Nie and H.~-Q.~Zhang,
  Phys.\ Rev.\ D {\bf 82} (2010) 066007
  [arXiv:1007.3321 [hep-th]].
  
\bibitem{Brihaye:2010mr}
  Y.~Brihaye and B.~Hartmann,
  Phys.\ Rev.\ D {\bf 81} (2010) 126008
  [arXiv:1003.5130 [hep-th]].

\bibitem{Barclay:2010up}
  L.~Barclay, R.~Gregory, S.~Kanno and P.~Sutcliffe,
  JHEP {\bf 1012} (2010) 029
  [arXiv:1009.1991 [hep-th]].

\bibitem{Jing:2010cx}
  J.~Jing, L.~Wang, Q.~Pan and S.~Chen,
  Phys.\ Rev.\ D {\bf 83} (2011) 066010
  [arXiv:1012.0644 [gr-qc]].

\bibitem{Gregory:2010yr}
  R.~Gregory,
  J.\ Phys.\ Conf.\ Ser.\  {\bf 283} (2011) 012016
  [arXiv:1012.1558 [hep-th]].

\bibitem{Barclay:2010nm}
  L.~Barclay,
  JHEP {\bf 1110} (2011) 044
  [arXiv:1012.3074 [hep-th]].

\bibitem{Li:2011xja}
  H.~-F.~Li, R.~-G.~Cai and H.~-Q.~Zhang,
  JHEP {\bf 1104} (2011) 028
  [arXiv:1103.2833 [hep-th]].

\bibitem{Kanno:2011cs}
  S.~Kanno,
  Class.\ Quant.\ Grav.\  {\bf 28} (2011) 127001
  [arXiv:1103.5022 [hep-th]].

\bibitem{Gangopadhyay:2012am}
  S.~Gangopadhyay and D.~Roychowdhury,
  JHEP {\bf 1205} (2012) 002
  [arXiv:1201.6520 [hep-th]].

\bibitem{Gangopadhyay:2012np}
  S.~Gangopadhyay and D.~Roychowdhury,
  JHEP {\bf 1205} (2012) 156
  [arXiv:1204.0673 [hep-th]].

\bibitem{Banerjee:2012vk}
  R.~Banerjee, S.~Gangopadhyay, D.~Roychowdhury and A.~Lala,
  Phys.\ Rev.\ D {\bf 87} (2013) 104001
  [arXiv:1208.5902 [hep-th]].

\bibitem{Momeni:2011iw}
  D.~Momeni, E.~Nakano, M.~R.~Setare and W.~-Y.~Wen,
  Int.\ J.\ Mod.\ Phys.\ A {\bf 28} (2013) 1350024
  [arXiv:1108.4340 [hep-th]].

\bibitem{Roychowdhury:2012hp}
  D.~Roychowdhury,
  Phys.\ Rev.\ D {\bf 86} (2012) 106009
  [arXiv:1211.0904 [hep-th]].

\bibitem{Franco:2009yz}
  S.~Franco, A.~Garcia-Garcia and D.~Rodriguez-Gomez,
  JHEP {\bf 1004} (2010) 092
  [arXiv:0906.1214 [hep-th]].
\bibitem{Horowitz:2010jq}
  G.~T.~Horowitz and B.~Way,
  JHEP {\bf 1011} (2010) 011
  [arXiv:1007.3714 [hep-th]].
\bibitem{Cai:2011ky}
  R.~-G.~Cai, H.~-F.~Li and H.~-Q.~Zhang,
  Phys.\ Rev.\ D {\bf 83} (2011) 126007
  [arXiv:1103.5568 [hep-th]].

\bibitem{Nakano:2008xc}
  E.~Nakano and W.~-Y.~Wen,
  Phys.\ Rev.\ D {\bf 78} (2008) 046004
  [arXiv:0804.3180 [hep-th]].

\bibitem{Albash:2008eh}
  T.~Albash and C.~V.~Johnson,
  JHEP {\bf 0809} (2008) 121
  [arXiv:0804.3466 [hep-th]].

\bibitem{Wen:2008pb}
  W.~-Y.~Wen,
  arXiv:0805.1550 [hep-th].

\bibitem{Albash:2009ix}
  T.~Albash and C.~V.~Johnson,
  arXiv:0906.0519 [hep-th].

\bibitem{Ge:2010aa}
  X.~-H.~Ge, B.~Wang, S.~-F.~Wu and G.~-H.~Yang,
  JHEP {\bf 1008} (2010) 108
  [arXiv:1002.4901 [hep-th]].

\bibitem{Ge:2011cw}
  X.~-H.~Ge,
  Prog.\ Theor.\ Phys.\  {\bf 128} (2012) 1211
  [arXiv:1105.4333 [hep-th]].
  
\bibitem{Domenech:2010nf}
  O.~Domenech, M.~Montull, A.~Pomarol, A.~Salvio and P.~J.~Silva,
  JHEP {\bf 1008} (2010) 033
  [arXiv:1005.1776 [hep-th]].
  
\bibitem{Roychowdhury:2012vj}
  D.~Roychowdhury,
  Phys.\ Lett.\ B {\bf 718} (2013) 1089
  [arXiv:1211.1612 [hep-th]].

\bibitem{Momeni:2011iw}
  D.~Momeni, E.~Nakano, M.~R.~Setare and W.~-Y.~Wen,
  Int.\ J.\ Mod.\ Phys.\ A {\bf 28} (2013) 1350024
  [arXiv:1108.4340 [hep-th]].

\bibitem{Roychowdhury:2012hp}
  D.~Roychowdhury,
  Phys.\ Rev.\ D {\bf 86} (2012) 106009
  [arXiv:1211.0904 [hep-th]].
\bibitem{poole2007} C.P. Poole, H.A. Farach and R.J. Creswick, Superconductivity, Academic Press, The
Netherlands (2007).


\end{thebibliography}

\end{document}